\documentclass[aps,preprint]{revtex4-1}

\pdfoutput=1

\usepackage[utf8]{inputenc}
\usepackage{lmodern}
\usepackage{amsmath}
\usepackage{amssymb}
\usepackage{graphicx}
\usepackage{xcolor}
\usepackage{color}
\usepackage{amssymb}
\usepackage{amsthm}
\usepackage{epstopdf} 
\usepackage{dcolumn}
\usepackage{amsfonts}
\usepackage{bm}

\newcommand{\cc}{\,\mathrm{cm^{-3}}}
\newcommand{\wcm}{\,\mathrm{W/cm}^2}
\newcommand{\mic}{\,\mu\mathrm{m}}
\newcommand{\nm}{\,\mathrm{nm}}
\newcommand{\mev}{\,\mathrm{MeV}}

\newcommand{\fs}{\,\mathrm{fs}}

\newcommand{\mrad}{\,\mathrm{mrad}}

\begin{document}

\title{Vacuum laser acceleration of relativistic electrons \\ using plasma mirror injectors}

\author{$^{1,*}$M. Thévenet, $^{2,*}$A. Leblanc, $^2$S. Kahaly, $^1$H. Vincenti, $^1$A. Vernier, $^2$F. Quéré and $^1$J. Faure}

\affiliation{$^1$LOA, ENSTA Paristech, CNRS, Ecole Polytechnique, Université Paris-Saclay, Palaiseau, France \\ 
$^2$LIDYL, DSM/IRAMIS, CEA Saclay, Gif-sur-Yvette, France}%

\begin{abstract}
Accelerating particles to relativistic energies over very short distances using lasers has been a long standing goal in physics. Among the various schemes proposed for electrons, vacuum laser acceleration has attracted considerable interest and has been extensively studied theoretically because of its appealing simplicity: electrons interact with an intense laser field in vacuum and can be continuously accelerated, provided they remain at a given phase of the field until they escape the laser beam. But demonstrating this effect experimentally has proved extremely challenging, as it imposes stringent requirements on the conditions of injection of electrons in the laser field. Here, we solve this long-standing experimental problem for the first time by using a plasma mirror to inject electrons in an ultraintense laser field, and obtain clear evidence of vacuum laser acceleration. With the advent of PetaWatt class lasers, this scheme could provide a competitive source of very high charge (nC) and ultrashort relativistic electron beams.
\end{abstract}

\maketitle

Femtosecond lasers currently achieve light intensities at focus that far exceed $10^{18}\,\mathrm{W/cm^2}$ at near infrared wavelengths \cite{yu12}. One of the great prospects of these extreme intensities is the laser-driven acceleration of electrons to relativistic energies within very short distances. At present, the most advanced scheme consists of using ultraintense laser pulses to excite large amplitude wakefields in underdense plasmas, providing extremely high accelerating gradients in the order of $100\,\mathrm{GV/m}$ \cite{esar09}. However, over the past decades, the direct acceleration of electrons by light \textit{in vacuum} has also attracted considerable interest and has been extensively studied theoretically \cite{hart95,esar95b,yu00,stu01,sal02,Pang2002, dodi03,malt03,varin06}. These investigations have been driven by the fundamental interest of this most elementary interaction, and by its potential for extreme electron acceleration through electric fields of $>10$’s TV/m that ultraintense laser pulses provide.

The underlying idea is to inject free electrons into an ultraintense laser field so that they always remain within a given half optical cycle of the field, where they constantly gain energy until they leave the focal volume. 1D Analytical calculations \cite{hart95} show that for relativistic electrons, the \textit{maximum} energy gain from this process is $\Delta E \propto mc^2\gamma_0 a_0^2$, where $\gamma_0$ is the electron initial Lorentz factor, and $a_0$ is the normalized laser vector potential, $m$ the electron mass, and $c$ the vacuum light velocity. Reaching high energy gains thus requires high initial energies $\gamma_0\gg1$ and/or ultrahigh laser amplitudes ($a_0 \gg 1$). 

In contrast with the large body of theoretical work published on this vacuum laser acceleration (VLA) of electrons to relativistic energies, experimental observations have largely remained elusive \cite{malk97b, mcna98,moor99, Payeur12, Cline13, Carbajo15} -sometimes even controversial \cite{ques98b, mcdo98}- and have so far not demonstrated significant energy gains. This is because VLA occurs efficiently only for electrons injected in the laser field with specific initial conditions that are extremely challenging to fulfill experimentally \cite{dodi03}. Indeed, in order to stay in phase with the laser field, electrons need to have initial velocities close to $c$ along the laser propagation axis. In addition, they should start interacting with the intense laser beam already close to its spatial and temporal maxima, and even be injected at appropriate phases of this field.

Electrons that do not satisfy these stringent requirements tend to explore many different optical cycles as they interact with the laser field, leading to an oscillatory motion where they are successively accelerated and decelerated, so that their final energy gain is low. When averaged over several cycles, this typically results in a drift motion where electrons are isotropically expelled away from high intensity regions, an effect which can be accounted for by the relativistic ponderomotive force \cite{schm73,star97,ques98}.  

Here, we present clear evidence of vacuum laser acceleration of bunches of $\sim 10^{10}$ electrons, corresponding to charges in the nC range, up to relativistic energies around 10 MeV. Our experimental results clearly discriminate for the first time electrons that have experienced a quasi-monotonic sub-laser-cycle acceleration, from those whose dynamics has mostly been determined by ponderomotive scattering. To solve the long-standing experimental problem of electron injection in the laser field, we demonstrate a new approach based on the use of \textit{plasma mirrors} \cite{MP-NP}, that specularly reflect ultraintense laser fields while simultaneously injecting relativistic electrons in the core of these reflected fields, co-linearly to the propagation direction. 


\subsection*{Plasma mirrors as electron injectors}

Plasma mirrors are dense plasmas resulting from the ionization of initially-solid targets irradiated by intense femtosecond laser pulses \cite{MP-NP}. Since their density is comparable to the initial solid density ($\approx10^{23}$ electrons$/\mathrm{cm}^{-3}$), their reflectivity can be as high as $80 \%$ \cite{doum04}. An essential feature of plasma mirrors is that to a large extent they behave like ordinary mirrors: the laser field is specularly reflected with hardly any alteration of its spatial properties \cite{MP-NC}, even at extremely high laser intensities. This is because plasma expansion is very limited on subpicosecond time scales: the plasma-vacuum interface remains optically flat (flatness $\delta L \ll \lambda$, with $\lambda$ the laser wavelength) while the femtosecond laser pulse reflects, thus leaving the beam wavefront essentially unaffected. Due to these remarkable properties, plasma mirrors are now largely used in ultrafast optics as a single-shot high-intensity optical device, e.g. to improve temporal contrast of femtosecond pulses \cite{doum04}, for the tight focusing of ultraintense beams \cite{PMfoc}, or for the generation of high-order harmonics and attosecond pulses \cite{Dromey,Lighthouse}.

\begin{figure*}[t!]
\centerline{\includegraphics[width=1\textwidth]{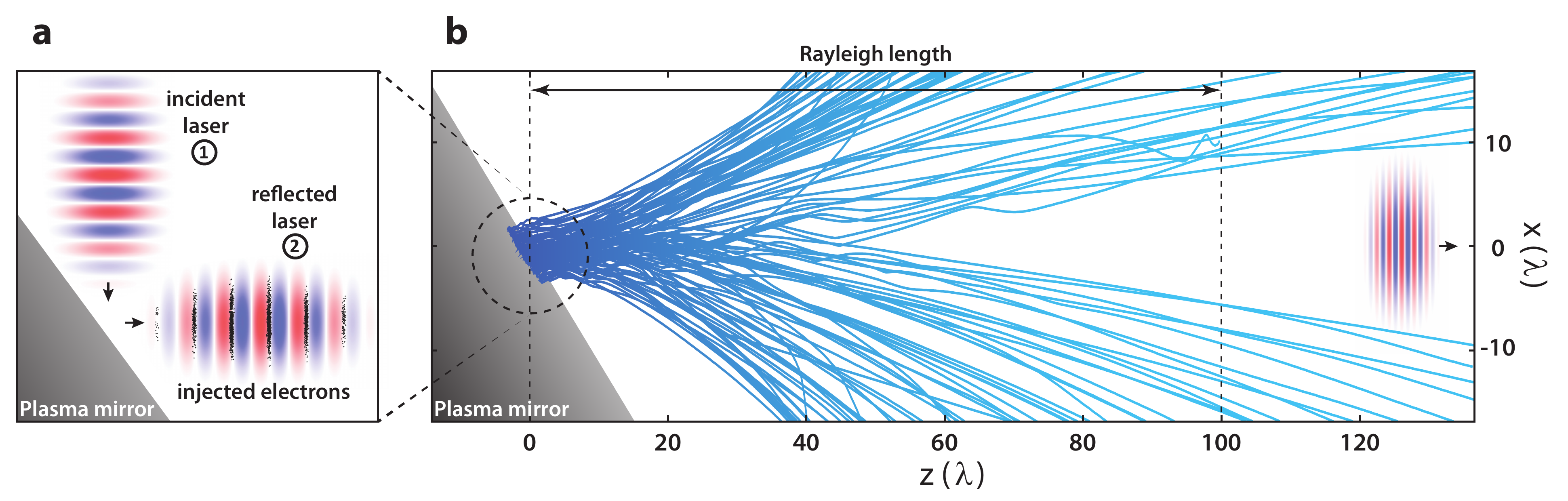}}
\caption{\textbf{Injection of relativistic electrons in ultraintense laser fields using plasma mirrors}. Panel a): principle of a plasma mirror injector. As an ultraintense laser pulse (E-field sketched in red and blue) reflects on a plasma mirror, it expels relativistic electrons (black dots) at specific phases of the field. These electrons then interact with the reflected pulse in vacuum. Panel b) shows electron trajectories (blue lines) computed from a 2D PIC simulation of the laser-plasma interaction (see Methods section). Electrons were initially located in the vicinity of the surface of the plasma mirror. Once expelled from the surface, they co-propagate and interact with the reflected laser field over a distance of the order of the Rayleigh length ($z_R=100\lambda$). This interaction clearly modifies the electron angular distribution as electrons are expelled to the side of the focal volume (see also the movie in the Suppl. Info.).}\label{fig1}
\end{figure*}

Upon reflection on a plasma mirror, an ultraintense laser pulse can also expel relativistic electrons in a direction close to the specular angle, as seen in Particle in Cell simulations (PIC) \cite{gein10}. Thus, the key idea of this work is that plasma mirrors can be used as electron injectors in the reflected laser field, providing a simple experimental solution to study the interaction of free electrons with intense lasers in vacuum (Fig.~\ref{fig1}). These electrons are emitted at given phases of the laser field \cite{gein10,tian12} as depicted in Fig.~\ref{fig1}a) and then interact with this field in vacuum over the Rayleigh length, see Fig.~\ref{fig1}b). Although there have been some observations of electron ejection from solid surfaces \cite{cai03,mordo09,wang10,tian12}, the experimental parameters did not permit to reach the VLA regime we have identified in this work, the intensity being too low ($<10^{19}$ $W/cm^2$) or the gradient scale length too long.

\subsection*{Experimental results}

Our experiment consisted in measuring the spatial profile of electron beams emitted by plasma mirrors exposed to ultraintense laser pulses, as well as their energy distributions at different emission angles. It was performed on the UHI100 laser of CEA/IRAMIS, a $100\,\mathrm{TW}$ laser system that delivers $800\nm$, $25\,\fs$ pulses. Once their temporal contrast is improved by 4 orders of magnitude with a plasma-based temporal filtering system \cite{DPM}, the laser pulses are focused on a flat fused silica target with an incidence angle of $55^\circ$ and in $p$-polarization, at a peak intensity of $2.10^{19}\,\wcm$ ($a_0 \simeq 3.1$). The focal spot is $5.5\,\mathrm{\mu m}$ at FWHM and the Rayleigh length is about $80\,\mathrm{\mu m}$. To optimize the electron signal, the scale length of the plasma density gradient at the surface was accurately controlled by pre-ionizing the target with a weaker prepulse ($\approx 10^{16}\,\wcm$) at an adjustable delay of a few hundreds femtoseconds before the main pulse \cite{kaha13} (see Suppl. Info. for the full data). The electron beam spatial profile was measured using a LANEX phosphor screen placed $15\,\mathrm{cm}$ away from the target, perpendicularly to the specular direction, imaged on a CCD camera. A magnetic spectrometer can be inserted before this LANEX screen in order to measure the electron energy distribution.

\begin{figure*}[t!]
\centerline{\includegraphics[width=.8\textwidth]{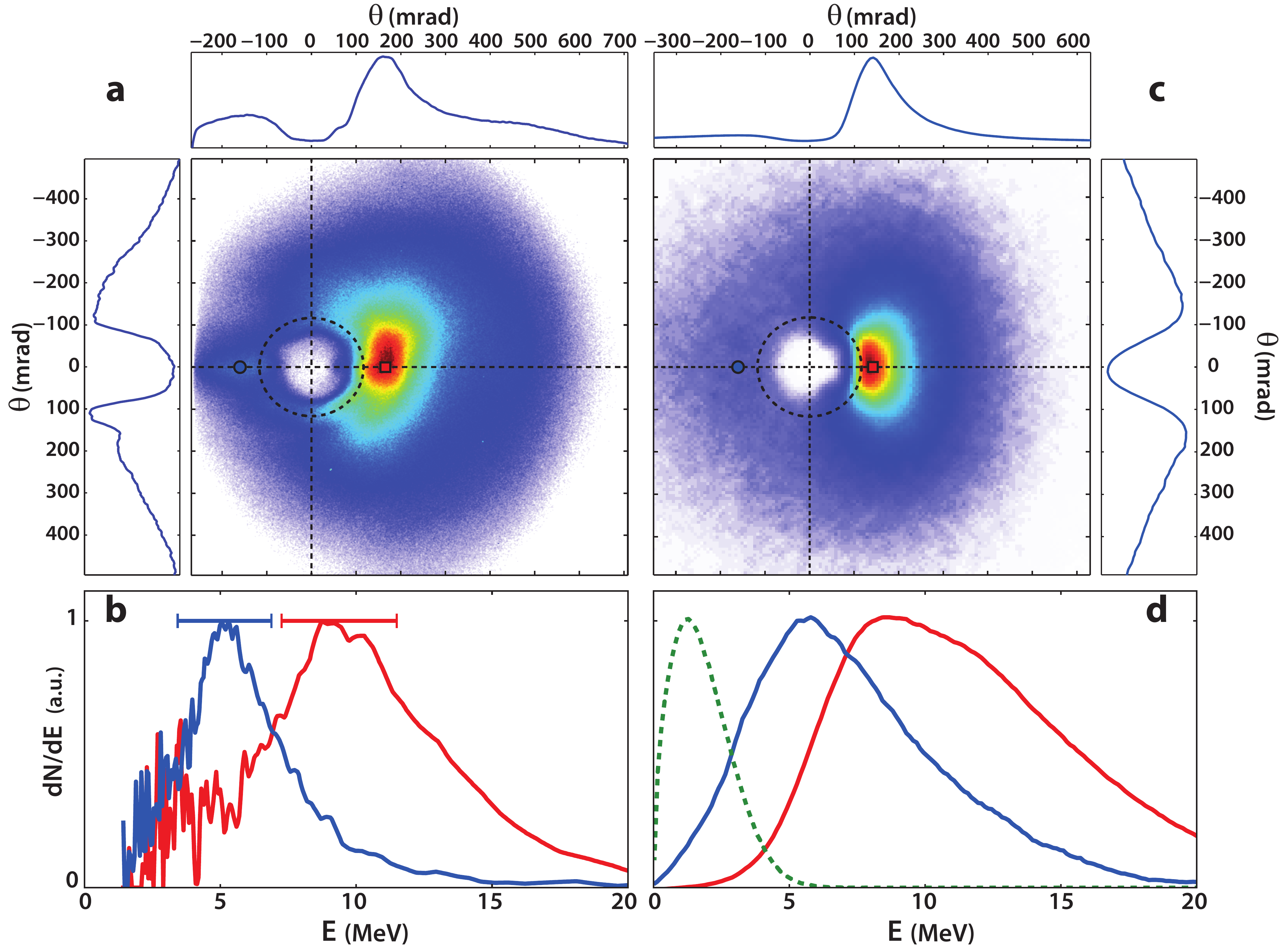}}
\caption{\textbf{Experimental evidence of vacuum laser acceleration.} Panel a) shows a typical experimental angular distribution of electrons emitted from plasma mirrors into the vacuum,  measured with the LANEX screen. It consists of a broad emission cone (blue disk), which is strongly modulated by two main patterns. One is a well defined hole (in white) around the reflected laser beam (whose size and position in the detection plane are indicated by the dashed circle), due to the ponderomotive scattering of  electrons after their ejection from the plasma mirror. The other is a bright peak (in red), right on the edge of this hole, due to Vacuum Laser Acceleration of a fraction of these electrons. Line-outs of the distribution along the dashed lines are plotted in the side panels, and the direction normal to the plasma mirror surface corresponds to $\theta \simeq 960\mrad$ (not shown). Panel b) shows the electron spectra measured at two different locations in the beam (the horizontal error bars represent the spectrometer resolution). These locations are indicated by the blue circle and the red square in panel a), that respectively correspond to the blue and red curves of panel b). All the features of panel a) and b) were very robust experimentally, being observed on all shots performed in similar experimental conditions (see Suppl. Info.). Panel c) and d) show the same quantities, now obtained from numerical simulations based on a 3D test particle model. The dashed curve in panel d) shows the initial electron energy distribution used in this model. }\label{fig2}
\end{figure*}

A typical electron angular distribution is shown in Fig.~\ref{fig2}. The measured electron emission spreads over a broad cone of $\approx600\,\mathrm{mrad}$ angular width, but is spatially very inhomogeneous within this cone. A pronounced hole is observed around the propagation direction of the reflected beam, with a cylindrical symmetry around this axis. The total angular width of this hole is about $200\,\mathrm{mrad}$, comparable to the divergence of the reflected beam. The other dominant feature is a bright electron peak on one edge of this hole, along the direction of the laser polarization (horizontal axis in Fig.~\ref{fig2}) and located between the specular and the normal direction. Its divergence is about $100\,\mathrm{mrad}$, much smaller than that of the total electron beam. From the signal measured on the calibrated LANEX screen \cite{glin06}, the overall charge in the electron beam is about $12\,\mathrm{nC}$, with $3\,\mathrm{nC}$ within the bright spot. Note that these patterns were clearly observed at high intensity $I>10^{19}\wcm$ and optimized for gradient scale lengths of $L\simeq\lambda/15$ (see Suppl. Info.).

Figure~\ref{fig2}b shows the spatially-resolved electron spectra measured at two different locations in the beam, on opposite sides of the hole along the laser polarization direction. Broad peaked spectra are observed, with a central energy of a few MeV. The key feature here is that the central energy is two times higher in the bright electron peak (10 MeV) than on the opposite side of the hole (5 MeV). This difference in energy suggests a straightforward interpretation for the spatial pattern of the electron beam. The bright peak along the laser polarization could correspond to electrons which have gained energy by VLA due to appropriate injection conditions in the field. The other electrons in the beam would have experienced isotropic ponderomotive scattering, leading to the symmetrical hole around the laser axis. We now validate this interpretation, by first studying the conditions of injection of electrons from plasma mirrors into the vacuum, and then their subsequent dynamics in the reflected field.

\subsection*{Modeling of the experimental results}

\begin{figure*}[t!]
\centerline{\includegraphics[width=.8\textwidth]{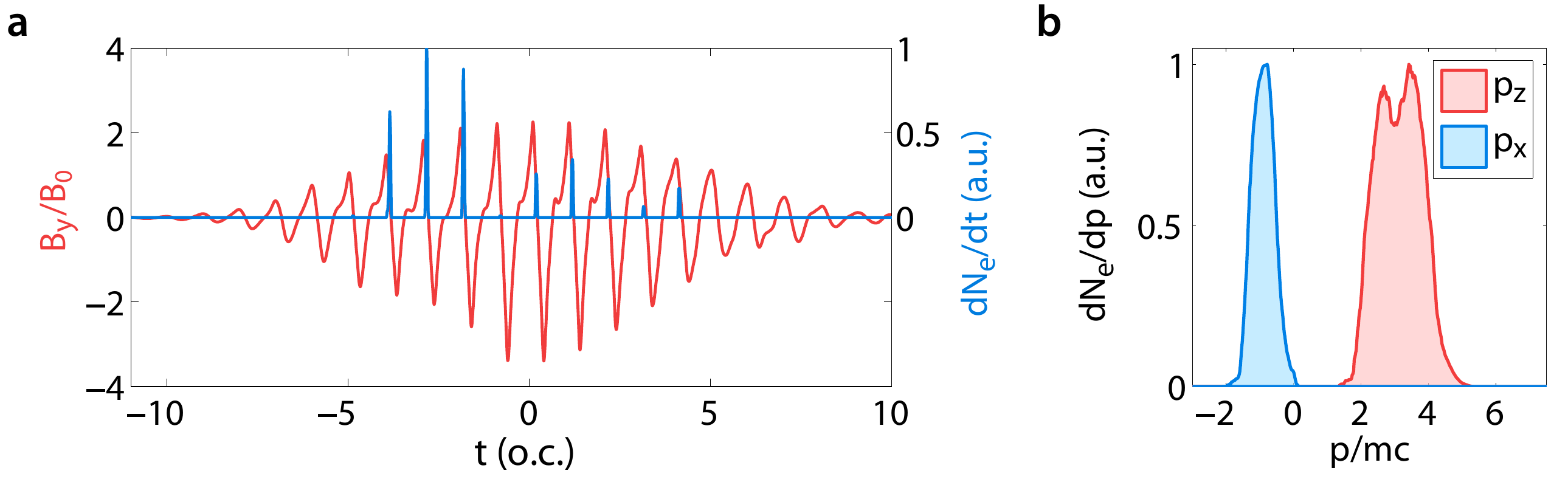}}
\caption{\textbf{Initial conditions of electrons ejected from plasma mirrors.} Panel a): waveform of the laser magnetic field $B_y$ reflected by the plasma mirror (red line), and temporal density profile of the ejected high-energy electrons (blue line), obtained from PIC simulations. Both quantities were sampled at $z=750\,\mathrm{nm}$ from the plasma mirror surface. Note that due to the complex coupling of the laser field with the plasma surface \cite{JP2}, the electron injection significantly drops around the peak of the laser pulse and most electrons are injected slightly before the pulse maximum. Panel b) shows the corresponding momentum distribution of these electrons, along the specular direction ($p_z$) and along the polarization direction of the reflected laser ($p_x$).}
\label{fig3}
\end{figure*}

Given the complexity of the coupling between intense laser fields and plasma mirrors, we turn to Particle In Cell (PIC) simulations to determine the properties of electrons ejected in the vacuum. Fig.~\ref{fig3}a) shows the density of ejected electrons as a function of time, together with the waveform $B_y(t)$ of the reflected field for interaction conditions corresponding to our experiment. These quantities were both sampled very close to the plasma mirror surface ($d\leq \lambda$), when electrons escape the plasma and are injected into the vacuum in the core of the reflected field. Electrons are observed to be emitted in the form of attosecond bunches at very precise phases close to the nodes of the laser field. We note that the waveform of the reflected field is distorted upon reflection, and strongly deviates from a pure sine wave. This is due to the generation of high-order harmonics of the laser frequency on the plasma mirror \cite{thau10}, which is unavoidable at these intensities in $p$-polarization. While this can quantitatively affect the exact outcome of the subsequent laser-electron interaction in vacuum, we will show that this does not qualitatively alter the involved physics. Figure~\ref{fig3}b) shows the momentum distributions of these electrons right after their ejection, along the specular direction ($p_z$) and along the laser polarization direction ($p_x$). Electrons start their motion in vacuum with relativistic velocities, corresponding to an average energy of $1.5\,\mathrm{MeV}$ ($\gamma_0\simeq 3$), and are ejected with an average angle of $20^\circ$ away from the specular direction. 

These initial conditions are close to being ideal for the observation of VLA, and definitely much more favorable than those achieved in all previous experimental attempts to observe this effect \cite{mcna98,moor99,Cline13, Carbajo15}. In experiments based on electron injection by ionization of core atomic levels, electrons started the interaction at rest and could not reach relativistic energies \cite{mcna98,moor99}. In those relying on electron beams produced by conventional accelerators, combined with an intense laser through a drilled mirror, the phase of injection in the field covers a full optical period, so that only a very small fraction of the electrons actually gains energy by VLA \cite{Cline13, Carbajo15}. 

To study the subsequent interaction of these electrons with the laser field in vacuum, we turn to a simple 3D test particle model, similar to the one used in \cite{ques98} (see Methods). In this model, the relativistic equations of motion are solved for electrons injected in a sinusoidal laser field, assumed to be Gaussian in space and time, and known analytically at every time and position. This is computationally much less demanding than the 3D PIC simulations that would be required to account for the isotropic effect of the ponderomotive force. 

Using this model, we calculate the trajectories of millions of electrons injected in the field. The set of initial conditions for these electrons is derived from the output of PIC simulations, such as those shown in Fig.~\ref{fig3} (see Methods). Figure~\ref{fig2}c) shows the angular electron distribution obtained from these simulations, for physical conditions corresponding to our experiment. The agreement of this distribution with the experimental one is striking: the two main features observed in the experiment -the hole around the laser axis and the bright peak along the laser polarization- are both well reproduced. The final energy spectra calculated on each side of the hole are shown in Fig.~\ref{fig2}d) and also compare well with the experimental observations. Despite its simplicity, this model thus captures the essential physics of the interaction in vacuum. This shows that effects such as space charge or the non-sinusoidal waveform of the reflected field, not taken into account in these simulations, do not play a major role once electrons are in vacuum.

\subsection*{Interpretation of the experimental results}

\begin{figure*}[t!]
\centerline{\includegraphics[width=.9\textwidth]{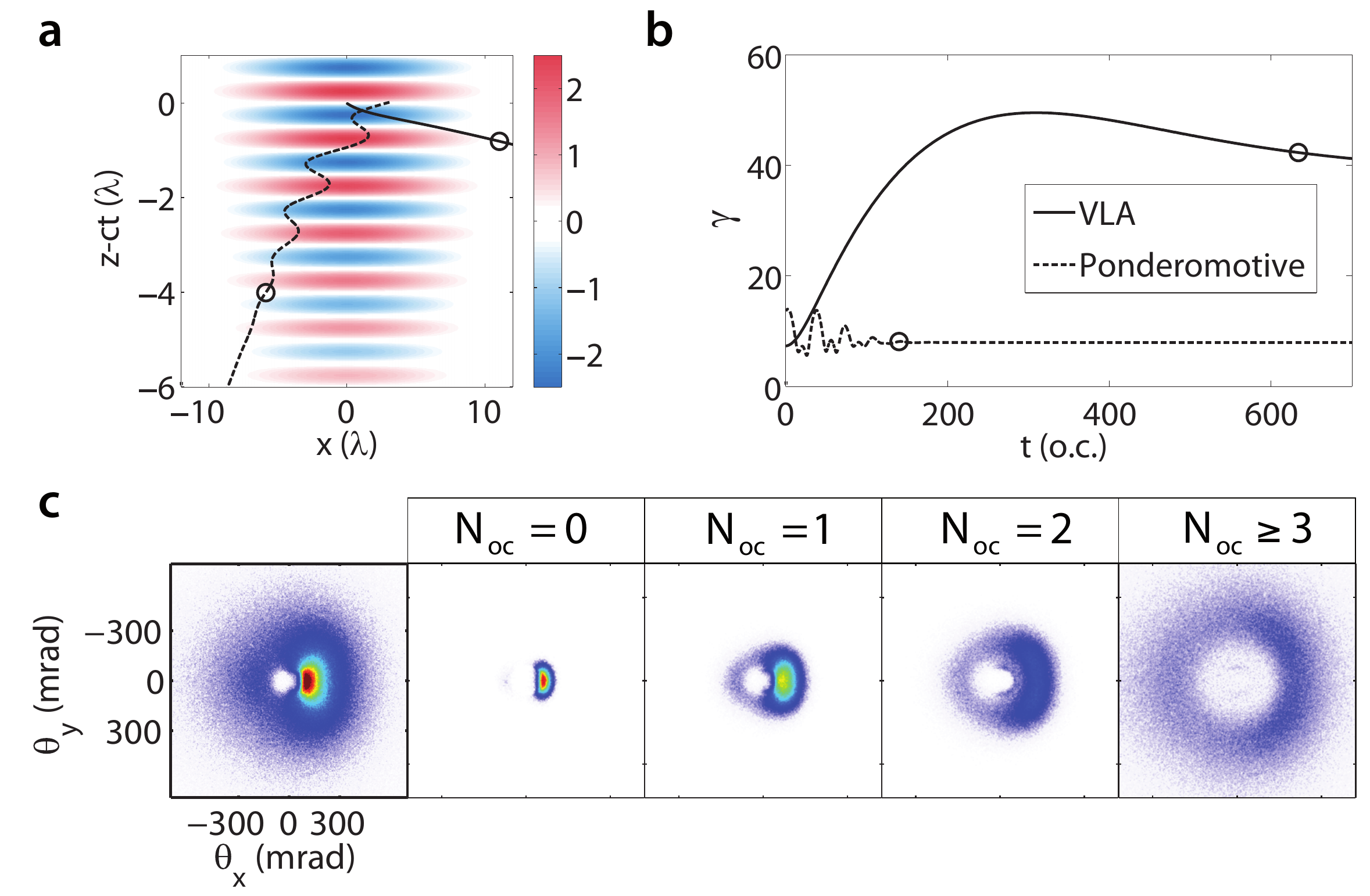}}
\caption{\textbf{3D modeling of laser-electron interaction in vacuum.} Panel a) shows two types of electron trajectories, corresponding to electrons that have respectively explored $N_{oc}\leq 1$ (full line) and $N_{oc}=3$ (dashed line) optical cycles before escaping the laser beam. The trajectories are displayed in a frame moving with the laser beam, and the color map sketches the laser field in this frame. The circles represent the positions for which we consider that these electrons escape the laser pulse (see Methods). Panel b) shows the temporal evolution of the electron Lorentz factor $\gamma(t)$ along these two trajectories. Panel c) shows the angular distribution for various electron populations which are sorted according to the number of optical cycles they experience. While VLA electrons clearly form the bright spot, ponderomotive electrons form a doughnut shape distribution. }
\label{fig4}
\end{figure*}

Considering this good agreement, this 3D model can now be exploited to analyze the trajectories of electrons contributing to the different patterns observed in the electron beam. To this end, we sort the electrons into different groups, depending on the number of laser optical cycles $N_{oc}$ that they have crossed along their trajectory. This provides a quantitative criterion for distinguishing electrons that experienced vacuum laser acceleration, from those that were scattered by the ponderomotive force. This is illustrated in Fig.~\ref{fig4}a) which shows two trajectories $(x,z-ct)$, representative of these two regimes: $N_{oc}\leq1$ (full line) and $N_{oc}=3$ (dashed line). The corresponding temporal evolutions of the electron Lorentz factor is shown in Fig.~\ref{fig4}b). 

In the first case, the ``VLA electron'' does not oscillate in the field, but rather ``surfs'' the laser wavefront in which it was injected, along the polarization direction. It thus gains energy almost all along its trajectory, until it escapes the focal volume sideways after an interaction distance of the order of the Rayleigh length. In contrast, the ``ponderomotive electron'' oscillates as it explores several optical cycles of the laser field, and gets quickly expelled out of the laser beam with a low energy gain -akin to a surfer that has missed the wave.  

In practice, there is a continuous transition between these two extreme types of trajectories, depending on the exact electron injection conditions in the field. This is illustrated by the images in Fig.~\ref{fig4}c) showing the beam patterns produced by several sub-ensembles of electrons in the simulations, corresponding to $N_{oc}$ varying from $N_{oc}=0$ (VLA electrons) to $N_{oc}\geq 3$ (ponderomotive electrons). Ponderomotive electrons form a doughnut-shaped beam centered on the laser propagation axis, while VLA electrons tend to concentrate in a bright peak on the edge of the ponderomotive hole, along the polarization direction. 

This analysis confirms our interpretation of the electron beam patterns observed in experiments, and provides clear proof that the bright peak in these patterns is due to VLA. According to our simulations, these electrons are accelerated from 1.5 MeV to 10 MeV (Fig. \ref{fig2}d)) over a distance of less than 100 $\mu m$, corresponding to an energy gain by VLA of about 7 in this experiment. In addition, a remarkable feature of VLA is that the position of the peak in Fig.~\ref{fig4}c) depends on the phase of injection of electrons in the laser field: if we artificially vary this phase by $\pi$ (half a laser period) in the simulation, this bright spot shifts to the other side of the ponderomotive hole. The experimental observation of this peak \textit{on one side only} of the hole is thus a clear indication that electrons are ejected out of the plasma mirror at a specific phase of the laser field, in the form of sub-laser cycle (attosecond) bunches.

\subsection*{Outlook}

We have obtained the first unambiguous experimental evidence of VLA of relativistic electrons by ultraintense laser fields, using a new concept for injecting electrons. This simple scheme, based on the remarkable properties of plasma mirrors, is flexible and can, in principle, be extended to much higher laser intensities and to more complex laser beams. From a fundamental point of view, it opens the way to the extensive experimental investigation of the interaction of free electrons with ultraintense lasers in various experimental conditions. Using PW-class lasers and intensities in excess of $10^{21}\wcm$, electrons could be exposed to accelerating gradients approaching 100 TV/m over tens of microns, and should thus reach energies in the GeV range \cite{sal02,malt03}, making VLA a promising scheme for the production of high charge (nC) ultra-relativistic beams. In addition, spatial shaping of the laser beam, providing doughnut beam shapes, as in radially polarized laser beams \cite{esar95b,stu01,varin06} or Laguerre-Gauss beams \cite{zhan08}, has the potential to further improve the beam quality and to provide more collimated electron beams. Similarly, temporal shaping of the field through the use of two-color lasers \cite{maur09,edwa14} could also be a path toward the sub-cycle control of the injection phase in the laser field.   

\section*{Methods}
\small{

\subsection*{PIC simulations}
2D Particle-In-Cell (PIC) simulations were carried out to gain insight into the injection and acceleration process (see Fig.~\ref{fig1}b) and movie in Suppl. Info.). A p-polarized laser pulse ($\lambda = 800\nm $, $\tau=25\fs \text{ FWHM}$, waist $w_0=5\mic$, $a_0=3.5$, where $a_0=e E_0/m \omega c $ is the normalized laser vector potential, with $e$ and $m$ the electron charge and mass, $\omega$ and $E_0$ the laser frequency and peak amplitude, and $c$ the vacuum light velocity) impinges on an overdense plasma ($n_{0}=100n_c$, where $n_c=1.6\times10^{21}\cc$ is the critical plasma density for $\lambda=800\nm$) with an exponential density gradient on its front side of decay length $L=\lambda/10$. The incidence angle is $45^\circ$. Simulation parameters were as follows: space step $\Delta x = \lambda/2000$, 40 particles per cell, box size: $40\lambda \times 45\lambda$. In order to follow the electrons trajectory far away from the target, a moving simulation box is used: after the laser pulse reflects from the target, the box begins to move at the speed of light and follows the reflected pulse, thus making it possible to follow the dynamics of energetic electrons along many wavelengths (typically $100 \mic$). \\
In order to determine the injection conditions of electrons in the reflected field,  we performed 1D PIC simulations in the boosted frame \cite{bour83}, with $\lambda = 800\nm$, $a_0 = 3$, $L = \lambda/8$, $\tau=25\fs$ FWHM. The numerical parameters were $\Delta x = \lambda/1000$, box length $4\lambda$, $4000$ particles per cell. The magnetic field and electron density plotted in Fig.3 were recorded at the front edge of the plasma mirror, at $z= 750\,\mathrm{nm}$ from the bulk density region (i.e. where $n=100n_c$). A filter was used to remove low-energy electrons ($E<0.5 \mev$), which typically return to the plasma within one optical cycle. Both 1D and 2D PIC simulations were performed with the code EPOCH. 

\subsection*{Model}
The 3D particle model consists in solving the relativistic equations of motion for electrons in a laser pulse. The Gaussian pulse propagates along the $+z$ direction and is polarized along $x$. The fields $E_x$ and $B_y$ are those given by the paraxial approximation. To properly model the ponderomotive force \cite{cicc90,ques98}, a first-order development with respect to the parameter $\epsilon = 1/kw_0$ is used in order to insure that the laser field satisfy Maxwell's equations, at least to the first order of $\epsilon$. This introduces new components $B_z$ and $E_z$, proportional to $\epsilon$ . A derivation of these fields can be found in Ref.~\cite{ques98}. Omitting these components leads to incorrect trajectories (as in \cite{malk97b,tian12}), by artificially restricting all forces to the polarization plane. We used PIC simulations to determine the reflectivity of the plasma mirror, which was about $70\%$ in our conditions. Therefore, we assumed that the amplitude of the reflected laser pulse was $a_0=2.5$ ($a_0=3$ for the incident field in the experiment). The other laser parameters were identical to the experimental ones: $\tau=25\fs$, $w_0=5\mic$, $\lambda=800\nm$.\\
The equations of motion were solved using a Boris pusher scheme, for $3\times10^6$ electrons.  For the initial conditions of motion, we used Gaussian distributions for all parameters (momentum components $p_x, p_y,p_z$ and emission time $t_0$). These distributions were all centered on the average values extracted from 1D PIC simulations described above. The widths of these distributions were used as adjustable parameters to reproduce the experimental data and obtain the simulations results of Fig.2. For all parameters, the widths that lead to the best agreement are found to be  larger than those provided by 1D PIC simulations. This difference can be attributed to two effects: 1- the actual variations of the interaction conditions (in particular the laser intensity) in 3D, and 2- the effect of the incident laser field on the electrons just after their emission. Indeed, the interference pattern of the incident and reflected beams affects the electron dynamics on a distance in the order of the beam waist $w_0$ (a few microns, which is small compared to the overall interaction length of electrons with the reflected field), and tends to significantly broaden the initial phase space distribution (see movie in Suppl. Info).\\
The calculation is run until all electrons have escaped the laser pulse which we consider to be true once the laser intensity at the electron position is $<1\%$ of its spatio-temporal maximum. For the sorting of electrons as a function of the number of optical cycles $N_{oc}$, we measure the electric field $E_x$ on electrons and detect the number of sign changes on $E_x$ along electron trajectories. This number, divided by $2$, is the number of optical cycles an electron has experienced.
}
\section*{Acknowledgements}
This work was funded by the European Research Council under Contract No. 306708, ERC Starting Grant FEMTOELEC and the Agence Nationale pour la Recherche under contract ANR-14-CE32-0011-03 APERO. We acknowledge the support of GENCI for access on super computer Curie. Simulations were run using EPOCH, which was developed as part of the UK EPSRC funded projects EP/G054940/1.

\section*{Author contributions}
*These two authors contributed equally to this work. \\
A.L. performed the experiment with S.K. and F.Q. A.L. analyzed the data, A.V. and J.F. calibrated the electron spectrometer. H.V. modified EPOCH for 1D boosted frame simulations. H.V. and M.T. performed the PIC simulations and developed the associated post-processing tools. M.T. developed and exploited the test particle model. All authors participated to the interpretation of the results. Figures were made by A.L. and M.T. F. Q. and J. F. designed and directed the project with equal contributions, and wrote the paper with inputs from the other authors.\\

\noindent
Competing Interests: The authors declare that they have no competing financial interests.\\

\noindent
Correspondence and requests for materials should be addressed to: jerome.faure@ensta.fr, fabien.quere@cea.fr


\newpage

\end{document}